\documentclass{article}

\begin{document}

\title{Born-Oppenheimer approximation for a harmonic molecule}
\author{Francisco M. Fern\'andez \thanks{%
Email: fernande@quimica.unlp.edu.ar} \\
INIFTA (UNLP, CONICET), Divisi\'{o}n Qu\'{i}mica Te\'{o}rica, \\
Diag. 113 y 64 (S/N), Sucursal 4, Casilla de Correo 16, 1900 La Plata,
Argentina}
\maketitle

\begin{abstract}
We apply the Born--Oppenheimer approximation to a harmonic
diatomic molecule with one electron. We compare the exact and
approximate results not only for the internal degrees of freedom
but also for the motion of the center of mass. We address the
problem of the permutation symmetry of identical nuclei and
discuss other applications of the model and its limitations.
\end{abstract}

\section{INTRODUCTION}

\label{sec:intro}

The first step in the treatment of a classical or quantum--mechanical
isolated system of particles should be the separation of the motion of the
center of mass from the internal degrees of freedom. Most textbooks on
quantum mechanics and quantum chemistry do that for the hydrogen atom but
then consider the nucleus at rest when they discuss many--electron atoms~%
\cite{CDL77,P68}. Such omission is more marked in the case of molecules
because they invariably resort to the Born--Oppenheimer (or clamped--nuclei)
approximation (BOA) and do not even mention the problem at all~\cite{P68}.
However, an adequate treatment of the motion of the center of mass is most
important for the estimation of adiabatic corrections to the
potential--energy hypersurface~\cite{K97}

Earlier pedagogical discussions of the BOA chose two--particle models: the
hydrogen atom~\cite{DM69} and a delta potential in a box~\cite{GD04}.
However, the simplest molecule, the hydrogen molecular ion $\mathrm{H}%
_{2}^{+}$, is a three--particle system. Therefore, those earlier discussions
are not sufficiently realistic to show many relevant features of the
treatment of molecular systems and of the nature of the BOA.

The purpose of this paper is the discussion of the separation of the motion
of the center of mass in the BOA. As already mentioned above, this aspect of
the problem is not discussed in most textbooks on quantum chemistry~\cite
{P68} and the BOA is almost entirely omitted from those on quantum mechanics~%
\cite{CDL77}. For simplicity we resort to a simple model of three particles
(two nuclei and one electron) that interact through Hooke's forces. Although
such harmonic interaction is unphysical, it has the advantage that the
Schr\"{o}dinger equation is solvable analytically and we can thus compare
the exact solution with the exact BO one.

In Section~\ref{sec:model} we present the model and obtain a dimensionless
Hamiltonian operator in a set of variables that allows the exact separation
of the motion of the center of mass. In Section~\ref{sec:exact} we solve the
Schr\"{o}dinger equation exactly. In Section~\ref{sec:BO} we obtain the
exact BO solution and compare it with an expansion of the exact nonadiabatic
result derived in Section~\ref{sec:exact}. In Section~\ref{sec:identical} we
address the interesting case of identical nuclei and discuss the
permutational symmetry of the molecular states and the correlation
functions. Finally, in Section~\ref{sec:comments} we discuss other
applications of the harmonic model and its limitations.

\section{MODEL}

\label{sec:model}

Our model consists of three particles of masses $M_{1}$, $M_{2}$ and $M_{3}$
that move in one dimension and interact through forces that follow Hooke's
law
\begin{eqnarray}
\hat{H} &=&-\frac{\hbar ^{2}}{2M_{1}}\frac{\partial ^{2}}{\partial X_{1}^{2}}%
-\frac{\hbar ^{2}}{2M_{2}}\frac{\partial ^{2}}{\partial X_{2}^{2}}-\frac{%
\hbar ^{2}}{2M_{3}}\frac{\partial ^{2}}{\partial X_{3}^{2}}  \nonumber \\
&&+\frac{1}{2}\left[ K_{12}\left( X_{1}-X_{2}\right) ^{2}+K_{13}\left(
X_{1}-X_{3}\right) ^{2}+K_{23}\left( X_{2}-X_{3}\right) ^{2}\right] ,
\label{eq:Ham_X}
\end{eqnarray}
where $K_{ij}$ are the force constants.

In order to make the Schr\"{o}dinger equation dimensionless we choose units
of length $L$, mass $M$ and force constant $K$ and define the corresponding
dimensionless quantities $x_{i}=X_{i}/L$, $m_{i}=M_{i}/M$ and $%
k_{ij}=K_{ij}/K$. If we choose $L=\left[ \hbar ^{2}/(MK)\right] ^{1/4}$ and
take into account that $\hbar ^{2}/\left( ML^{2}\right) =\hbar \omega $,
where $\omega =\sqrt{K/M}$, we easily derive the dimensionless Hamiltonian
operator
\begin{eqnarray}
\frac{\hat{H}}{\hbar \omega } &=&-\frac{1}{2m_{1}}\frac{\partial ^{2}}{%
\partial x_{1}^{2}}-\frac{1}{2m_{2}}\frac{\partial ^{2}}{\partial x_{2}^{2}}-%
\frac{1}{2m_{3}}\frac{\partial ^{2}}{\partial x_{3}^{2}}  \nonumber \\
&&+\frac{1}{2}\left[ k_{12}\left( x_{1}-x_{2}\right) ^{2}+k_{13}\left(
x_{1}-x_{3}\right) ^{2}+k_{23}\left( x_{2}-x_{3}\right) ^{2}\right] .
\label{eq:Ham_x}
\end{eqnarray}
From now on we write $\hat{H}$ instead of $\hat{H}/(\hbar \omega )$ and
simply remember that the energy is measured in units of $\hbar \omega $.
Note that we may choose $M=M_{i}$ and/or $K=K_{ij}$ in order to make some
particular parameters equal to unity and simplify the Schr\"{o}dinger
equation even further. However, we have decided to keep all the model
parameters for clarity.

In order to separate the motion of the center of mass from the internal
degrees of freedom we define new coordinates as follows:
\begin{eqnarray}
q_{1} &=&\frac{1}{m_{T}}\left( m_{1}x_{1}+m_{2}x_{2}+m_{3}x_{3}\right)
,\,m_{T}=m_{1}+m_{2}+m_{3},  \nonumber \\
q_{2} &=&x_{2}-x_{1},  \nonumber \\
q_{3} &=&x_{3}-x_{1}.  \label{eq:q(x)}
\end{eqnarray}
Note that $q_{1}$ is the coordinate of the center of mass and $q_{2}$ and $%
q_{3}$ are simply the positions of the particles 2 and 3 with respect to
particle 1. The latter variables are translationally invariant as it is
expected for the internal (spectroscopic) degrees of freedom. More
precisely, the displacement of the whole system $x_{i}\rightarrow x_{i}+a$
produces $q_{1}\rightarrow q_{1}+a$, $q_{2}\rightarrow q_{2}$ and $%
q_{3}\rightarrow q_{3}$. The Hamiltonian operator (\ref{eq:Ham_x}) (in units
of $\hbar \omega $) becomes
\begin{eqnarray}
\hat{H} &=&-\frac{1}{2m_{T}}\frac{\partial ^{2}}{\partial q_{1}^{2}}-\frac{1%
}{2m_{2}}\frac{\partial ^{2}}{\partial q_{2}^{2}}-\frac{1}{2m_{3}}\frac{%
\partial ^{2}}{\partial q_{3}^{2}}  \nonumber \\
&&-\frac{1}{2m_{1}}\left( \frac{\partial ^{2}}{\partial q_{1}^{2}}+\frac{%
\partial ^{2}}{\partial q_{3}^{2}}+2\frac{\partial ^{2}}{\partial
q_{2}\partial q_{3}}\right)   \nonumber \\
&&+\frac{1}{2}\left[ k_{12}q_{2}^{2}+k_{13}q_{3}^{2}+k_{23}\left(
q_{2}-q_{3}\right) ^{2}\right] .  \label{eq:Ham_q}
\end{eqnarray}

\section{EXACT SOLUTION}

\label{sec:exact}

The Hamiltonian operator (\ref{eq:Ham_q}) is the sum of an operator that
depends only on $q_{1}$%
\begin{equation}
\hat{H}_{1}=-\frac{1}{2m_{T}}\frac{\partial ^{2}}{\partial q_{1}^{2}},
\label{eq:H1}
\end{equation}
and another one that depends only on $q_{2}$ and $q_{3}$
\begin{eqnarray}
\hat{H}_{2} &=&-\frac{1}{2m_{2}}\frac{\partial ^{2}}{\partial q_{2}^{2}}-%
\frac{1}{2m_{3}}\frac{\partial ^{2}}{\partial q_{3}^{2}}-\frac{1}{2m_{1}}%
\left( \frac{\partial ^{2}}{\partial q_{1}^{2}}+\frac{\partial ^{2}}{%
\partial q_{3}^{2}}+2\frac{\partial ^{2}}{\partial q_{2}\partial q_{3}}%
\right)   \nonumber \\
&&+\frac{1}{2}\left[ k_{12}q_{2}^{2}+k_{13}q_{3}^{2}+k_{23}\left(
q_{2}-q_{3}\right) ^{2}\right] .  \label{eq:H23}
\end{eqnarray}
Therefore, each eigenfunction of $\hat{H}$ can be written as the product of
an eigenfunction of $\hat{H}_{1}$ times an eigenfunction of $\hat{H}_{2}$.
The operator $\hat{H}_{1}$ describes the motion of a free pseudoparticle
with mass equal to the total mass of the system. If $\Phi (q_{2},q_{3})$ is
an eigenfunction of $\hat{H}_{2}$
\begin{equation}
\hat{H}_{2}\Phi =\epsilon \Phi ,  \label{eq:Schr_23}
\end{equation}
we conclude that the eigenfunctions $\psi (q_{1},q_{2},q_{3})$ of $\hat{H}$
\begin{equation}
\hat{H}\psi =E\psi ,  \label{eq:Schrodinger}
\end{equation}
are of the form
\begin{equation}
\psi (q_{1},q_{2},q_{3})=e^{i\kappa q_{1}}\Phi (q_{2},q_{3}),  \label{eq:psi}
\end{equation}
and
\begin{equation}
E=\frac{\kappa ^{2}}{2m_{T}}+\epsilon ,  \label{eq:E}
\end{equation}
where $-\infty <\kappa <\infty $.

Note that the Hamiltonian operator $\hat{H}$ has a continuous spectrum, and
that the Hamiltonian for the internal degrees of freedom $\hat{H}_{2}$ has a
discrete (or point) one (which we will obtain below). This point is most
important when one has to calculate expectation values of observables. For
example, everybody knows that if the potential--energy function is
homogeneous of degree two (that is to say, it satisfies $V(t\mathbf{r}%
)=t^{2}V(\mathbf{r})$) then the virial theorem gives us $<\hat{T}>=<V>$ for
an eigenfunction of the Hamiltonian operator. However, this relationship
does not apply to the kinetic and potential energies in $\hat{H}$ because
the integrals diverge, but it applies to the kinetic and potential energies
in $\hat{H}_{2}$ since its eigenfunctions are square integrable.

The Hamiltonian operator (\ref{eq:H23}) is a particular case of
\begin{equation}
\hat{H}=-\frac{1}{2}\sum_{i}\sum_{j}A_{ij}\frac{\partial ^{2}}{\partial
q_{i}\partial q_{j}}+\frac{1}{2}\sum_{i}\sum_{j}B_{ij}q_{i}q_{j}.
\label{eq:H_bilin}
\end{equation}
In order to obtain its eigenfunctions and eigenvalues we carry out a change
of variables of the form
\begin{equation}
q_{i}=\sum_{j}c_{ij}y_{j}.  \label{eq:q(y)}
\end{equation}
Thus the Hamiltonian operator (\ref{eq:H_bilin}) becomes
\begin{equation}
\hat{H}=-\frac{1}{2}\sum_{i}\sum_{j}\left[ \mathbf{C}^{-1}\mathbf{A}\left(
\mathbf{C}^{-1}\right) ^{T}\right] _{ij}\frac{\partial ^{2}}{\partial
y_{i}\partial y_{j}}+\frac{1}{2}\sum_{i}\sum_{j}\left( \mathbf{C}^{T}\mathbf{%
BC}\right) _{ij}y_{i}y_{j},
\end{equation}
where $\mathbf{A}$, $\mathbf{B}$, and $\mathbf{C}$ are matrices with
elements $A_{ij}$, $B_{ij}$, and $c_{ij}$, respectively. We choose the
matrix $\mathbf{C}$ in such a way that
\begin{eqnarray}
\mathbf{C}^{-1}\mathbf{A}\left( \mathbf{C}^{-1}\right) ^{T} &=&\mathbf{I},
\nonumber \\
\mathbf{C}^{T}\mathbf{BC} &=&\mathbf{D},  \label{eq:diag_D}
\end{eqnarray}
where $\mathbf{D}$ is a diagonal matrix
\begin{equation}
\mathbf{D}_{ij}=\omega _{i}^{2}\delta _{ij}.
\end{equation}
This approach is well known in the treatment of small oscillations in
classical mechanics~\cite{G80}.

The resulting Hamiltonian operator is a sum of uncoupled dimensionless
harmonic oscillators
\begin{equation}
\hat{H}=-\frac{1}{2}\sum_{i}\frac{\partial ^{2}}{\partial y_{i}^{2}}+\frac{1%
}{2}\sum_{i}\omega _{i}^{2}y_{i}^{2},
\end{equation}
therefore its eigenfunctions are products
\begin{equation}
\Phi _{\{n\}}=\prod_{i}\varphi _{n_{i}}(y_{i}),
\end{equation}
and its eigenvalues are given by
\begin{equation}
\epsilon _{\{n\}}=\sum_{i}\epsilon _{n_{i}},\,\epsilon _{n_{i}}=\left( n_{i}+%
\frac{1}{2}\right) \omega _{i},  \label{eq:E_bilinear}
\end{equation}
where $n_{i}=0,1,\ldots $ are harmonic--oscillator quantum numbers, and each
$\omega _{i}$ is a dimensionless frequency. More precisely,
\begin{equation}
\left( -\frac{1}{2}\frac{\partial ^{2}}{\partial y_{i}^{2}}+\frac{1}{2}%
\omega _{i}^{2}y_{i}^{2}\right) \varphi _{n_{i}}(y_{i})=\epsilon
_{n_{i}}\varphi _{n_{i}}(y_{i}).
\end{equation}

It follows from equation (\ref{eq:diag_D}) that
\begin{equation}
\mathbf{C}^{-1}\mathbf{ABC}=\mathbf{D},  \label{eq:diag_D_2}
\end{equation}
which shows that the problem reduces to the diagonalization of the
nonsymmetric matrix $\mathbf{AB}$.

In the particular case of the Hamiltonian operator (\ref{eq:H23}) we have
\begin{eqnarray}
\mathbf{A} &=&\left(
\begin{array}{cc}
\frac{1}{m_{1}}+\frac{1}{m_{2}} & \frac{1}{m_{1}} \\
\frac{1}{m_{1}} & \frac{1}{m_{1}}+\frac{1}{m_{3}}
\end{array}
\right) ,  \nonumber \\
\mathbf{B} &=&\left(
\begin{array}{cc}
k_{12}+k_{23} & -k_{23} \\
-k_{23} & k_{12}+k_{23}
\end{array}
\right) ,  \nonumber \\
\mathbf{AB} &=&\left(
\begin{array}{cc}
\frac{k_{12}(m_{1}+m_{2})+k_{23}m_{1}}{m_{1}m_{2}} & \frac{%
k_{13}m_{2}-k_{23}m_{1}}{m_{1}m_{2}} \\
\frac{k_{12}m_{3}-k_{23}m_{1}}{m_{1}m_{3}} & \frac{%
k_{13}(m_{1}+m_{3})+k_{23}m_{1}}{m_{1}m_{3}}
\end{array}
\right) .  \label{eq:mat_A_B_AB}
\end{eqnarray}
The characteristic polynomial for the matrix $\mathbf{AB}$ is
\begin{eqnarray}
&&w^{2}-w\frac{%
k_{12}m_{3}(m_{1}+m_{2})+k_{13}m_{2}(m_{1}+m_{3})+k_{23}m_{1}(m_{2}+m_{3})}{%
m_{1}m_{2}m_{3}}  \nonumber \\
&&+\frac{(m_{1}+m_{2}+m_{3})\left[ k_{12}(k_{13}+k_{23})+k_{13}k_{23}\right]
}{m_{1}m_{2}m_{3}},  \label{eq:charpoly}
\end{eqnarray}
where $w=\omega ^{2}$. The two real positive roots give us the frequencies
that we need to obtain the energy eigenvalues according to equation (\ref
{eq:E_bilinear}):
\begin{equation}
\epsilon _{n_{1},n_{2}}=\omega _{1}\left( n_{1}+\frac{1}{2}\right) +\omega
_{2}\left( n_{2}+\frac{1}{2}\right) .  \label{eq:epsilon_n1_n2}
\end{equation}
We do not show those roots explicitly here because they are rather
cumbersome.

The exact eigenfunctions for the internal degrees of freedom
\begin{equation}
\Phi _{n_{1},n_{2}}(q_{2},q_{3})=\varphi _{n_{1}}(y_{2})\varphi
_{n_{2}}(y_{3}),  \label{eq:Phi(q2,q3)}
\end{equation}
clearly show the coupling of the motion of the particles through the
variables $y_{i}$ that are linear combinations of the $q_{j}$. In other
words, the problem is completely separable in the variables $y_{i}$ but not
in the $q_{j}$ or $x_{k}$.

\section{THE BORN--OPPENHEIMER APPROXIMATION}

\label{sec:BO}

The BOA is discussed in many textbooks~\cite{P68} and also in pedagogical
articles\cite{DM69,GD04}. For this reason we do not develop it here
explicitly and just show its results for the present model. From now on we
assume that our three--particle system models a diatomic molecule with just
one electron. We choose the particles 1 and 2 to be the nuclei and particle
3 to be the electron; more precisely, we assume that $m_{1}\geq m_{2}\gg
m_{3}$. We clearly appreciate that the harmonic potential chosen here is not
realistic because it describes an attractive interaction between the nuclei.
However, since the BOA is based on the different particle masses and not on
the nature of the interaction we can apply it successfully and compare its
approximate solutions with the exact ones.

In the clamped--nuclei approximation we omit the kinetic energy of the
nuclei, which we assume to be at rest at $x_{1}$ and $x_{2}$, and solve the
Schr\"{o}dinger equation for the remaining ``electronic'' Hamiltonian~\cite
{P68}
\begin{equation}
\hat{H}_{e}=-\frac{1}{2m_{3}}\frac{\partial ^{2}}{\partial x_{3}^{2}}+\frac{1%
}{2}\left[ k_{13}\left( x_{1}-x_{3}\right) ^{2}+k_{23}\left(
x_{2}-x_{3}\right) ^{2}\right] .  \label{eq:He}
\end{equation}
Since the ``internuclear'' interaction $k_{12}\left( x_{1}-x_{2}\right)
^{2}/2$ is just a constant we add it later to the electronic eigenvalues.

If we rewrite the potential--energy function as
\begin{eqnarray*}
&&\frac{1}{2}\left[ k_{13}\left( x_{1}-x_{3}\right) ^{2}+k_{23}\left(
x_{2}-x_{3}\right) ^{2}\right] =\frac{1}{2}\left( k_{13}+k_{23}\right)
\left[ x_{3}-\frac{k_{13}x_{1}+k_{23}x_{2}}{k_{13}+k_{23}}\right] ^{2} \\
&&-\frac{\left( k_{13}x_{1}+k_{23}x_{2}\right) ^{2}}{2\left(
k_{13}+k_{23}\right) }+\frac{1}{2}\left(
k_{13}x_{1}^{2}+k_{23}x_{2}^{2}\right) ,
\end{eqnarray*}
then we realize that the electronic Hamiltonian (\ref{eq:He}) is just a
displaced harmonic oscillator and that the electronic energies are given by
\begin{equation}
\epsilon _{e,n_{1}}(x_{1},x_{2})=\sqrt{\frac{k_{13}+k_{23}}{m_{3}}}\left(
n_{1}+\frac{1}{2}\right) -\frac{\left( k_{13}x_{1}+k_{23}x_{2}\right) ^{2}}{%
2\left( k_{13}+k_{23}\right) }+\frac{1}{2}\left(
k_{13}x_{1}^{2}+k_{23}x_{2}^{2}\right) .  \label{eq:E_e}
\end{equation}

The nuclear motion is governed by the potential--energy function
\begin{eqnarray}
U(x_{1},x_{2}) &=&U(x_{1}-x_{2})=\epsilon _{e,n_{1}}(x_{1},x_{2})+\frac{1}{2}%
k_{12}\left( x_{1}-x_{2}\right) ^{2}  \nonumber \\
&=&\sqrt{\frac{k_{13}+k_{23}}{m_{3}}}\left( n_{1}+\frac{1}{2}\right) +\frac{%
k_{13}k_{23}+k_{12}k_{13}+k_{12}k_{23}}{2\left( k_{13}+k_{23}\right) }\left(
x_{1}-x_{2}\right) ^{2}.  \nonumber \\
&&  \label{eq:U(x1-x2)}
\end{eqnarray}
The final step is the solution of the Schr\"{o}dinger equation for the
nuclear Hamiltonian operator
\begin{equation}
\hat{H}_{N}=-\frac{1}{2m_{1}}\frac{\partial ^{2}}{\partial x_{1}^{2}}-\frac{1%
}{2m_{2}}\frac{\partial ^{2}}{\partial x_{2}^{2}}+U(x_{1}-x_{2}).
\label{eq:H_N}
\end{equation}
In doing so, we separate the motion of the center of mass by means of the
change of variables
\begin{eqnarray}
R &=&\frac{1}{m_{N}}\left( m_{1}x_{1}+m_{2}x_{2}\right) ,\,m_{N}=m_{1}+m_{2}
\nonumber \\
q_{2} &=&x_{2}-x_{1}  \label{eq:R(x)}
\end{eqnarray}
and rewrite the Hamiltonian operator (\ref{eq:H_N}) as
\begin{equation}
\hat{H}_{N}=-\frac{1}{2m_{N}}\frac{\partial ^{2}}{\partial R^{2}}-\frac{1}{%
2\mu _{N}}\frac{\partial ^{2}}{\partial q_{2}^{2}}+U(q_{2}),\,\mu _{N}=\frac{%
m_{1}m_{2}}{m_{N}}.  \label{eq:H_N_2}
\end{equation}
The eigenfunctions of this operator are of the form
\begin{equation}
\Phi _{N}(R,q_{2})=e^{i\kappa R}\varphi (q_{2}),  \label{eq:Phi_N}
\end{equation}
and its eigenvalues are approximations to those of $\hat{H}$:
\begin{equation}
E_{\kappa ,n_{1},n_{2}}^{BO}=\frac{\kappa ^{2}}{2m_{N}}+\sqrt{\frac{%
k_{13}+k_{23}}{m_{3}}}\left( n_{1}+\frac{1}{2}\right) +\sqrt{\frac{%
k_{13}k_{23}+k_{12}k_{13}+k_{12}k_{23}}{\left( k_{13}+k_{23}\right) \mu _{N}}%
}\left( n_{2}+\frac{1}{2}\right) .  \label{eq:E^BO}
\end{equation}

If $\Phi _{e}(x_{3};x_{1},x_{2})$ is an eigenfunction of the electronic
Hamiltonian $\hat{H}_{e}$ then the full BOA eigenfunction is
\begin{equation}
\psi ^{BO}(q_{1},q_{2},q_{3})=e^{i\kappa R}\Phi
_{e}(x_{3};x_{1},x_{2})\varphi (q_{2}).  \label{eq:psi^BO}
\end{equation}
Although the BOA treats the electronic and nuclear motions separately, we
appreciate that the BOA solution (\ref{eq:psi^BO}) already couples the
nuclear and electronic degrees of freedom through the electronic function $%
\Phi _{e}(x_{3};x_{1},x_{2})$ much in the way the exact solution (\ref
{eq:Phi(q2,q3)}) does. Besides, it is worth noting that the BOA correctly
describes the internal degrees of freedom in terms of
translationally--invariant coordinates: $q_{2}$ and $%
x_{3}-(k_{13}x_{1}+k_{23}x_{2})/(k_{13}+k_{23})$.

Finally, we show that we can obtain the BO eigenvalues from the exact
solution (\ref{eq:epsilon_n1_n2}). Because the BOA is based on the fact that
the nuclear masses are much greater than the electronic ones, we substitute $%
m_{1}=u_{1}/\lambda $ and $m_{2}=u_{2}/\lambda $ into the roots $w_{1}$ and $%
w_{2}$ of the characteristic polynomial (\ref{eq:charpoly}) and rearrange
the results in order to remove the poles (for example, multiplying numerator
and denominator by $\lambda ^{2}$). If the undetermined quantities $u_{1}$
and $u_{2}$ are of the same order of magnitude as $m_{3}$ then $\lambda \ll 1
$. Then we expand both roots in Taylor series about $\lambda =0$: $%
w_{i}=w_{i,0}+w_{i,1}\lambda +\ldots $. After tedious algebraic manipulation
of the equations (greatly facilitated by any available computer algebra
software) we obtain
\begin{eqnarray}
w_{1} &=&\frac{k_{13}+k_{23}}{m_{3}}+\frac{k_{13}^{2}u_{2}+k_{23}^{2}u_{1}}{%
u_{1}u_{2}(k_{13}+k_{23})}\lambda +\ldots   \nonumber \\
&=&\frac{k_{13}+k_{23}}{m_{3}}+\frac{k_{13}^{2}m_{2}+k_{23}^{2}m_{1}}{%
m_{1}m_{2}(k_{13}+k_{23})}+\ldots ,  \nonumber \\
w_{2} &=&\frac{(u_{1}+u_{2})\left[ k_{12}(k_{13}+k_{23})+k_{13}k_{23}\right]
}{u_{1}u_{2}(k_{13}+k_{23})}\lambda +\ldots   \nonumber \\
&=&\frac{(m_{1}+m_{2})\left( k_{12}k_{13}+k_{12}k_{23}+k_{13}k_{23}\right) }{%
m_{1}m_{2}(k_{13}+k_{23})}+\ldots .  \label{eq:w1,w2}
\end{eqnarray}
The first contribution to $w_{1}$ gives the large frequency due to the fast
motion of the electron (second term in the right--hand side of equation (\ref
{eq:E^BO})), and the dominant contribution to $w_{2}$ provides the frequency
for the slow motion of the nuclei (third term in the right--hand side of
equation (\ref{eq:E^BO})). If, in addition, we take into account that $%
m_{T}\approx m_{N}$ and $q_{1}\approx R$ when we neglect the electron mass,
we clearly appreciate that the BOA yields the leading terms of the expansion
of the exact result in negative powers of the nuclear masses (or the ratio
of the electron to nuclear mass), even for the contribution of the motion of
the center of mass. We have decided to express that ratio conveniently as $%
\lambda =u_{i}/m_{i}$ but other forms are possible.

An alternative mathematical strategy for obtaining the expansions of the
roots of the characteristic polynomial is to substitute $w=w^{(0)}+w^{(1)}%
\lambda +\ldots $ (and, of course, $m_{i}=u_{i}/\lambda $) into the
characteristic polynomial (\ref{eq:charpoly}) and expand the resulting
expression in powers of $\lambda $. At order zero we obtain two roots $%
w_{1}^{(0)}\neq 0$ and $w_{2}^{(0)}=0$, and the coefficients of greater
powers of $\lambda $ yield further corrections. The result should also be
equation (\ref{eq:w1,w2}).

\section{IDENTICAL NUCLEI}

\label{sec:identical}

The particular case of identical nuclei leads to simpler expressions because
\begin{equation}
m_{1}=m_{2},\,k_{13}=k_{23}.
\end{equation}
We easily obtain
\begin{equation}
\omega _{1}^{2}=\frac{k_{13}(2m_{1}+m_{3})}{m_{1}m_{3}},\,\omega _{2}^{2}=%
\frac{2k_{12}+k_{13}}{m_{1}},
\end{equation}
and
\begin{equation}
y_{2}=\frac{\sqrt{2m_{1}m_{3}}(2q_{3}-q_{2})}{2\sqrt{2m_{1}+m_{3}}},\,y_{3}=%
\frac{\sqrt{2m_{1}}}{2}q_{2}.
\end{equation}

We appreciate that $\hat{P}_{12}q_{1}=q_{1}$, $\hat{P}_{12}q_{2}=-q_{2}$ and
$\hat{P}_{12}q_{3}=q_{3}-q_{2}$, where $\hat{P}_{12}$ is the permutation
operator that satisfies $\hat{P}_{12}f(x_{1},x_{2})=f(x_{2},x_{1})$.
Therefore, $\hat{P}_{12}y_{2}=y_{2}$ and $\hat{P}_{12}y_{3}=-y_{3}$, so that
we conclude that
\begin{equation}
\hat{P}_{12}\psi (x_{1},x_{2},x_{3})=(-1)^{n_{2}}\psi (x_{1},x_{2},x_{3}).
\end{equation}
In this way we can build symmetric and antisymmetric states (including the
spin) for boson and fermions, respectively.

At first sight it may seem that when choosing the particle 1 as the
coordinate origin we are violating the quantum--mechanical principle that
identical particles are indistinguishable. However, when we express the
resulting wavefunctions in terms of the original variables we realize that
we can take into account the correct permutational symmetry explicitly, and,
therefore, there is no violation of that principle. That the identical
particles are treated exactly in the same way is more clearly seen in the
form of the variables that appear in the exact square--integrable
eigenfunctions: $q_{2}=x_{2}-x_{1}$ and $2q_{3}-q_{2}=2x_{3}-(x_{1}+x_{2})/2$%
. We also realize that the BO states exhibit exactly the same symmetry as
follows from the fact that the nuclear factor is a function of $q_{2}$ and $%
\Phi _{e}(x_{3};x_{1},x_{2})$ actually depends on $2x_{3}-(x_{1}+x_{2})/2$.

The eigenfunctions are somewhat complicated to write them down explicitly
here. However, the correlation functions for the ground state are not so
cumbersome. For example,
\begin{equation}
\rho (x_{1},x_{2})=\int_{-\infty }^{\infty }\Phi
_{0,0}(q_{2},q_{3})^{2}\,dx_{3}=\frac{\sqrt{m_{1}\omega _{2}}}{\sqrt{2\pi }}%
e^{-\frac{m_{1}\omega _{2}}{2}\left( x_{1}-x_{2}\right) ^{2}},
\label{eq:rho(x1,x2)}
\end{equation}
gives us the probability of finding one nucleus at $x_{2}$ if the other one
is at $x_{1}$. The fact that it exhibits a maximum at $x_{1}=x_{2}$ is a
consequence of the unrealistic attractive internuclear interaction of our
model. This equation clearly shows that we are correctly treating both
nuclei as indistinguishable particles.

If, on the other hand, we integrate over the coordinate of one of the nuclei
we have
\begin{eqnarray}
\rho (x_{1},x_{3}) &=&\int_{-\infty }^{\infty }\Phi
_{0,0}(q_{2},q_{3})^{2}\,dx_{2}  \nonumber \\
&=&\frac{\sqrt{2m_{1}m_{3}\omega _{1}\omega _{2}}}{\sqrt{\pi \left[
2m_{1}\omega _{2}+m_{3}(\omega _{1}+\omega _{2})\right] }}e^{-\frac{%
2m_{1}m_{3}\omega _{1}\omega _{2}}{2m_{1}\omega _{2}+m_{3}(\omega
_{1}+\omega _{2})}(x_{1}-x_{3})^{2}},  \label{eq:rho(x1,x3)}
\end{eqnarray}
that shows the coupling between the electronic and nuclear motions.

The BOA yields remarkably similar expressions: $\rho ^{BO}(x_{1},x_{2})=\rho
(x_{1},x_{2})$ and
\begin{equation}
\rho ^{BO}(x_{1},x_{3})=\frac{\sqrt{2m_{1}m_{3}\omega _{1}\omega _{2}}}{%
\sqrt{\pi \left( 2m_{1}\omega _{2}+m_{3}\omega _{1}\right) }}e^{-\frac{%
2m_{1}m_{3}\omega _{1}\omega _{2}}{2m_{1}\omega _{2}+m_{3}\omega _{1}}%
(x_{1}-x_{3})^{2}}.  \label{eq:rho^BO(x1,x3)}
\end{equation}
Note that the only difference between the exact (\ref{eq:rho(x1,x3)}) and BO
(\ref{eq:rho^BO(x1,x3)}) correlation functions is the neglect of the small
frequency with respect to the large one: $\omega _{1}+\omega _{2}\approx
\omega _{1}$. We appreciate that for most purposes the BOA gives a correct
description of the system behavior, at least according to the simple
harmonic model discussed here.

\section{FURTHER COMMENTS AND CONCLUSIONS}

\label{sec:comments}

In this paper we have applied the BOA to an exactly solvable model for a
diatomic molecule with one electron. We could thus show that the
clamped--nuclei approach provides the leading terms of the expansion of the
eigenfunctions and eigenvalues in powers of the ratio of the electron to the
nuclear mass. In addition to the internal degrees of freedom we also
considered the motion of the center of mass that is commonly omitted in most
textbooks on quantum mechanics and quantum chemistry~\cite{CDL77,P68}.
Although molecular physicists are more interested in the internal
(spectroscopic) degrees of freedom we explicitly considered the factor that
corresponds to the motion of the center of mass for completeness. As
discussed above for the virial theorem, only the internal degrees of freedom
should be taken into account for the calculation of expectation values and
transition probabilities. The reader may find a rigorous discussion of the
treatment of the motion of the center of mass for actual molecular systems
elsewhere~\cite{K97}.

The harmonic potential--energy function of the present model is unrealistic,
but such a choice allows us to solve the Schr\"{o}dinger equation
analytically and thus compare the exact solution with the BO one. Unlike
earlier oversimplified pedagogical models used to discuss the BOA~\cite
{DM69,GD04}, our harmonic molecule contains the minimum number of particles
to be a molecule and is therefore more realistic from this point of view.
For example, we could address the interesting case of identical nuclei.

The harmonic model is not suitable for the discussion of the Franck--Condon
principle~\cite{C47} because all the electronic curves exhibit exactly the
same nuclear equilibrium distance ($q_{2}=0$) and frequency as shown by
equation (\ref{eq:U(x1-x2)}). Therefore, the value of an overlap integral is
either one or zero because the harmonic--oscillator eigenfunctions are
orthogonal.

The three--particle harmonic model is also useful for the study of the mass
polarization in atoms~\cite{BD03}. In fact, if we choose the particles 2 and
3 to be electrons we have a one--dimensional harmonic version of the Helium
atom. We can thus, for example, estimate the effect of neglecting the motion
of the nucleus, or discuss the isotope effects on the atomic properties.

Our discussion of the harmonic diatomic molecule may also serve as an
introduction to the non Born--Oppenheimer calculation of molecular
properties~\cite{CA02b} that requires an adequate separation of the motion
of the center of mass in the way shown in Sec.~\ref{sec:model}.

\section*{Acknowledgment}

The author would like to thank Professor J. F. Ogilvie for useful comments
and suggestions.

\end{document}